\documentclass{article}

\usepackage{algorithm}
\usepackage{algpseudocode}
\usepackage{authblk}
\usepackage{amssymb}
\usepackage{amsmath}
\usepackage{amsthm}
\usepackage{array}
\usepackage{booktabs}
\usepackage{cleveref} 
\usepackage{geometry}
\usepackage{graphicx}
\usepackage{longtable}
\usepackage{multirow}
\usepackage{multicol}
\usepackage{subcaption}
\usepackage{tikz}

\usetikzlibrary{calc}

\crefname{figure}{Figure}{Figures}
\crefname{table}{Table}{Tables}

\begin{document}

\title{Boosting Rectilinear Steiner Minimum Tree Algorithms with Augmented Bounding Volume Hierarchy}
\author{Puhan Yang, Guchan Li \thanks{Equal contribution. Email: yph21@mails.tsinghua.edu.cn, li-gc22@mails.tsinghua.edu.cn}}
\affil{Tsinghua University}

\date{}

\maketitle

\begin{abstract}
    The rectilinear Steiner minimum tree (RSMT) problem computes the shortest network connecting a given set of points using only horizontal and vertical lines, possibly adding extra points (Steiner points) to minimize the total length. RSMT solvers seek to balance speed and accuracy. In this work, we design a framework to boost existing RSMT solvers, extending the Pareto front. Combined with GeoSteiner, our algorithm reaches 5.16\% length error on nets with 1000 pins. The average time needed is 0.46 seconds. This provides an effective way to solve large-scale RSMT problems with small-scale solvers.
\end{abstract}

\section{Introduction}
The Rectilinear Steiner Minimum Tree (RSMT) problem is a fundamental problem in combinatorial optimization and computational geometry. It is widely applied in electronic design automation (EDA), network topology design, and logistics planning. The goal is to construct a Steiner tree with the shortest total rectilinear (Manhattan) distance by adding a minimal number of additional Steiner points. Since the RSMT problem is NP-hard, various heuristic and approximation methods have been developed to balance computational efficiency and solution quality.

\subsection{Classical RSMT Algorithms}
In this section, we examine traditional RSMT solvers. These solvers focus on approximating solutions using minimum spanning tree, dynamic spanning tree, and local optimization. 

One classical approach is the Kou-Markowsky-Berman (KMB) algorithm \cite{kou1981fast}, which is based on minimum spanning tree. The algorithm first computes the minimum spanning tree of the given terminal set, then introduces Manhattan median points as candidate Steiner points, and finally reconnects edges to form a Steiner tree with reduced total length. This method has a time complexity of $O(n \log n)$ and guarantees a worst-case approximation ratio of 2.

Another significant method is Zelikovsky’s approximation algorithm  \cite{zelikovsky1993approximations}, which improves MST-based methods by iteratively adding additional Steiner points to further reduce path lengths and using a local optimization strategy to enhance the tree structure. This method achieves a theoretical approximation ratio of $11/6$ ($\approx 1.833$) and has a computational complexity of $O(n^2)$, making it more accurate than KMB but computationally expensive.

Hwang’s algorithm \cite{hwang1976steiner} leverages grid-based rectilinear structures and applies dynamic programming to Steiner point selection. Although it offers a tighter approximation, its high computational cost limits its use in large-scale applications.

\subsection{Modern High-Efficiency RSMT Algorithms}
With the increasing complexity of VLSI designs, modern algorithms prioritize both accuracy and computational efficiency. Several methods have gained widespread adoption due to their scalability and performance.

FLUTE (Fast Lookup Table-based Estimation of Steiner Trees) is the most widely used RSMT algorithm in VLSI physical design \cite{chu2008flute}. It achieves remarkable efficiency by precomputing and storing optimal Steiner trees for terminal sets of less than 9 points in a lookup table, allowing O(1) query time for small-scale RSMT solutions. For larger terminal sets, it employs a recursive partitioning heuristic. FLUTE is an industry standard for RSMT computation due to its exceptional speed and near-optimal results.

BOI (Batched Oblivious Improvement) is an efficient heuristic method that improves RSMT construction incrementally \cite{griffith1994steiner}. It generates an initial Steiner tree and applies iterative local refinements to reduce total path length, stopping once no further improvements are possible. With a linear time complexity of O(n), BOI is particularly effective for large-scale RSMT problems.

FastSteiner combines greedy heuristics and local optimization to refine Steiner point placement \cite{hwang1992steiner}. Its key advantages include handling obstacle-aware RSMT computation, unlike FLUTE, achieving an approximation ratio of 1.1 to 1.2 for high accuracy, and being well-suited for FPGA routing and irregular grid layouts. FastSteiner is widely applied in scenarios requiring obstacle-aware routing, such as FPGA placement and high-density circuit layouts.

GeoSteiner is a dynamic programming and branch-and-bound-based algorithm originally designed for Euclidean Steiner Trees (EST) but adapted for RSMT computations \cite{warme2000geosteiner}. It computes exact optimal solutions for small instances and employs branch-and-bound techniques for larger instances.

BFLUTE (Batched FLUTE) is an enhanced version of FLUTE designed for batch processing in large-scale VLSI layout optimization \cite{wong2008bflute}. It processes multiple RSMT instances simultaneously, reducing redundant computations and making it suitable for full-chip routing optimization rather than just local interconnections.

\subsection{REST}
REST (Rectilinear Edge Sequence Tree) \cite{liu2021rest} is a deep reinforcement learning based algorithm for constructing the RSMT. The core innovation in REST lies in its actor-critic algorithm that iteratively refines RSMT solutions. Unlike traditional methods that rely on predefined heuristics, REST learns to construct near-optimal solutions through experience. REST achieves an average length error of less than 0.36\% for nets with up to 50 terminals. REST utilizes modern GPUs to accelerate computation. In terms of performance, REST is significantly faster than FLUTE and BGA while maintaining a solution close to optimal.

However, the convergence rate of REST is low. It might take considerable time and hardware resources to train an agent for each degree. Although the network structure allows REST to compute RSMT for any graph, it suffers from distribution shift between training and testing graphs. This limits the ability of REST to solve large problems. In our experiment, directly applying an agent that is trained on 50-terminal graphs to 1000-terminal graphs leads to an error rate of 50\%. Although more advanced deep RL algorithms can be developed, our focus is to directly boost existing REST agents.

\subsection{Conclusion}
We divide RMST solvers into three categories according to speed and accuracy. Class 1 algorithms have both high speed and accuracy. FLUTE is a typical class 1 algorithm. Class 2 algorithms have high accuracy, but with higher asymptotic complexity. A typical example for class 2 algorithm is GeoSteiner. Class 3 algorithms have low accuracy or scalability, but have high potential. We chose REST as a representative for class 3 algorithms.

\section{Algorithm}
We aim to boost class 2 and class 3 RSMT solvers into class 1 solvers. Our algorithm runs in $O(n \log n)$ time regardless of underlying solvers. It not only extended the Pareto front of RSMT solvers, but also fixed issues of existing solvers. Specifically, augmented GeoSteiner successfully surpassed all tested algorithms in both speed and accuracy. The error rate of REST reduced from 53.4\% to 15.4\%. Accuracy level of FLUTE was maintained.

\subsection{Motivation}
We divide and conquer RSMT problem by splitting the graph into independent blocks. The chosen RSMT solver computes subtree in each block. All subtrees are treated as single nodes, which are then connected by minimum spanning tree. We use L1 norm to ensure rectilinearity. Let $N$ be the number of points, and $B$ be the maximum block size.

% partitioned graph
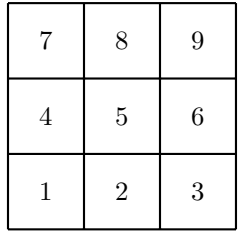
\begin{figure}[ht]
    \centering
    \begin{tikzpicture}
        \draw[step=1cm, thick] (0,0) grid (3,3);

        \foreach \x in {0,1,2} {
          \foreach \y in {0,1,2} {
            \node at (\x+0.5, \y+0.5) {\pgfmathparse{int(3*\y + \x + 1)}\pgfmathresult};
          }
        }
    \end{tikzpicture}
    \caption{Partitioned Graph}
    \label{fig:divide}
\end{figure}

A possible solution is dividing the graph with rectangular grids, as shown in \cref{fig:divide}. Blocks 1 and 9 are not likely to be connected in miminum spanning tree, thus the links between them can be erased. Furthermore, only $O(N)$ edges are needed to compute global minimum spanning tree. Information about neighbors can be maintained using fast multipole method (FMM) \cite{greengard1987fast}. However, this simple dividing strategy might lead to suboptimal results.

To better partition the graph, we use bounding volume hierarchy (BVH), which is locality sensitive. However, we are unable to compute neighbors for each block. Computing minimum spanning tree for a graph with $O(N^2)$ edges requires $O(N^2 \log N)$ time. 

It would be desirable to combine FMM and BVH together. In the following sections, we augment BVH to partition the graph wisely, while maintaining neighboring information.

\subsection{Segment}
Intuitively speaking, a block can lookup its neighbors by checking its edges. If each edge stores its neighboring blocks, then the problem is solved. Problem arises when an edge is adjacent to multiple blocks. It is not straightforward to maintain a list in each edge with low cost.

% define segment
\begin{figure}[ht]
    \centering
    \begin{tikzpicture}[scale=1]
        \draw[thick] (-2, 0) -- (0, 0);
        \coordinate (midpoint1) at (-1, 0);
        \draw[->, thick] (midpoint1) -- ($(midpoint1) + (0, 0.5)$);
        \draw[->, thick] (midpoint1) -- ($(midpoint1) + (0, -0.5)$);
        \node[above, font=\large] at ($(midpoint1) + (0, 0.6)$) {front};
        \node[below, font=\large] at ($(midpoint1) + (0, -0.6)$) {back};

        \draw[thick] (2, -1) -- (2, 1);
    
        \coordinate (midpoint2) at (2, 0);
    
        \draw[->, thick] (midpoint2) -- ($(midpoint2) + (0.5, 0)$);
        \draw[->, thick] (midpoint2) -- ($(midpoint2) + (-0.5, 0)$);
    
        \node[right, font=\large] at ($(midpoint2) + (0.6, 0)$) {front};
        \node[left, font=\large] at ($(midpoint2) + (-0.6, 0)$) {back};
    \end{tikzpicture}
    \caption{Segment}
    \label{fig:segment}
\end{figure}

We intruduce the notion of segment, which is the maximal subsegment of an edge where it is adjacent to only two blocks. Each block maintains a list of segments that surround itself. As shown in \cref{fig:segment}, a segment can access its neighboring blocks using $front$ and $back$ pointers. Each node (block) stores a list called $segments$.

\subsection{Construction}
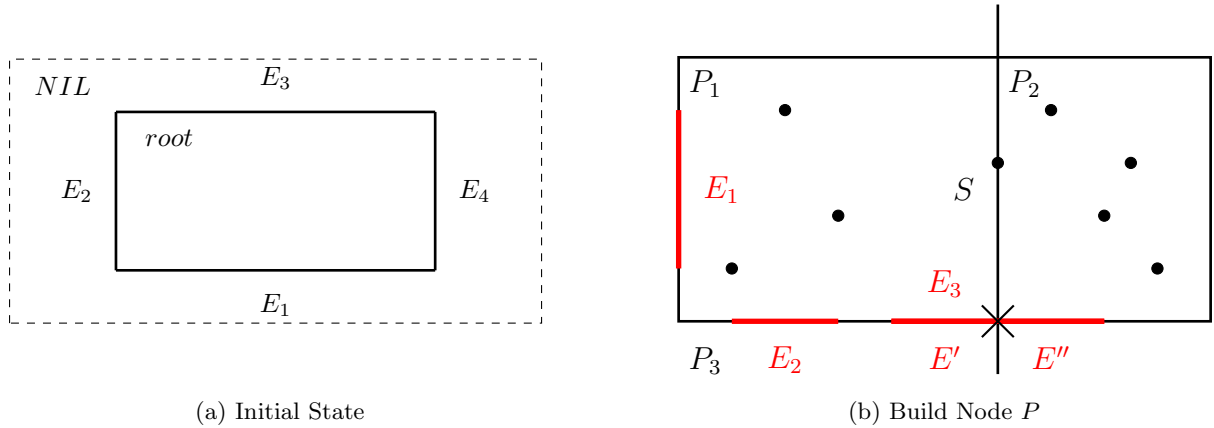
\begin{figure}[htbp]
    \centering
    \begin{subfigure}[b]{0.48\textwidth}
        \centering
        \begin{tikzpicture}[scale=0.7]
            \node at (-4, 2) {$NIL$};
            \node at (-2, 1) {$root$};
    
            \draw[dashed] (-5, -2.5) rectangle (5, 2.5);
            
            \draw[line width=1pt] (-3, -1.5) -- (3, -1.5) node [midway, below, yshift=-5pt] {$E_1$};
            \draw[line width=1pt] (-3, -1.5) -- (-3, 1.5) node [midway, left, xshift=-5pt] {$E_2$};
            \draw[line width=1pt] (-3, 1.5) -- (3, 1.5) node [midway, above, yshift=5pt] {$E_3$};
            \draw[line width=1pt] (3, -1.5) -- (3, 1.5) node [midway, right, xshift=5pt] {$E_4$};
        \end{tikzpicture}
        \vspace{22pt}
        \caption{Initial State}
        \label{fig:init}
    \end{subfigure}
    \hfill
    \begin{subfigure}[b]{0.48\textwidth}
        \centering
        \begin{tikzpicture}[scale=0.7]
            \draw[line width=1pt] (0, 0) rectangle (10, 5); 
    
            \node[font=\large] at (0.5, 4.5) {$P_1$};
            \node[font=\large] at (6.5, 4.5) {$P_2$};
            \node[font=\large] at (0.5, -0.75) {$P_3$};
        
            \draw[line width=1pt] (6, -1) -- (6, 6) node [midway, left, xshift=-5pt, font=\large] {$S$}; 
    
            \draw[line width=2pt, red] (0, 1) -- (0, 4) node [midway, right, xshift=5pt, font=\large] {$E_1$};
    
            \draw[line width=2pt, red] (1, 0) -- (3, 0) node [midway, below, yshift=-5pt, font=\large] {$E_2$}; 
        
            \draw[line width=2pt, red] (4, 0) -- (8, 0) node [pos=0.25, above, yshift=5pt, font=\large] {$E_3$}; 
    
            \draw[line width=2pt, red] (4, 0) -- (6, 0) node [midway, below, yshift=-5pt, font=\large] {$E'$};
    
            \draw[line width=2pt, red] (6, 0) -- (8, 0) node [midway, below, yshift=-5pt, font=\large] {$E''$};
            
            % draw some points
            \filldraw[black] (1, 1) circle (3pt);
            \filldraw[black] (2, 4) circle (3pt);
            \filldraw[black] (3, 2) circle (3pt);
            \filldraw[black] (6, 3) circle (3pt);
            \filldraw[black] (9, 1) circle (3pt);
            \filldraw[black] (8, 2) circle (3pt);
            \filldraw[black] (8.5, 3) circle (3pt);
            \filldraw[black] (7, 4) circle (3pt);
    
            % cross point
            \draw[thick] (5.7, -0.3) -- (6.3, 0.3);
            \draw[thick] (5.7, 0.3) -- (6.3, -0.3);
        \end{tikzpicture}
        \caption{Build Node $P$}
        \label{fig:build} 
    \end{subfigure}
    \caption{Constructing BVH}
    \label{fig:construct}
\end{figure}

Nodes in ordinary BVH are augmented to maintain segments. \cref{fig:init} shows the initial state of our algorithm. Root node of BVH consists all points. The virtual node NIL is used to replace null pointers. Initially, all segments are the edges of root's bounding box.

BVH is built recursively from root to leaf. Each node inside BVH corresponds to a block in the graph. Each node splits along the longest dimension of its bounding box. This procedure includes first sorting all points, then distributing points evenly into two child nodes. Recursion terminates when the number of points $\le B$.

In augmented BVH, information about segments are maintained during the recursion. In \cref{fig:build}, $P$ splits horizontally into node $P_1$ and $P_2$ with $S$ as cutting line. Each segment $E$ in $P.segments$ is updated. There are three possible cases:

\begin{itemize}
    \item $E_1$ is parallel to $S$. $E.front$ is set to $P_1$, and $E$ is added into $P_1.segments$.
    \item $E_2$ is vertical to $S$ but does not intersect $S$. The operations are the same with case 1.
    \item $E_3$ intersects $S$. It is no longer a segment and splits into $E'$ and $E''$, which are added into $P_1.segments$ and $P_2.segments$ respectively. $P_3=E_3.back$ also replaces $E$ with $E'$ and $E''$ in its segment list. Finally, the front and back pointers of $E'$ and $E''$ are set correctly to $P_1$, $P_2$ and $P_3$.
\end{itemize}

The last step is to create a new segment for $S$. We first set $S.back=P_1$ and $S.front=P_2$, then add $S$ into $P_1.segments$ and $P_2.segments$. After visiting $P$, $P.segments$ can be cleared to save memory. This leads to the fact that all segments are stored in leaf nodes and NIL.

\subsection{Analysis}
In this section we analyze the time and space complexity for constructing augmented BVH. For $N$ input points, augmented BVH can be constructed in $O(N\log N)$ time using $O(N)$ space. The total time complexity equals the cost for constructing ordinary BVH plus the cost introduced by segments.

\subsubsection{Time Complexity for Constructing BVH (updated)}
Assume the total time for building ordinary BVH is $T_1(n)$ with $n$ points. From previous discussions, we have
$$T_1(n) = 2 \times T_1(\frac{n}{2}) + O(n\log n)$$
According to Master Theorem, $T_1(n) = O(n\log n)$. The total time for constructing BVH is $O(N\log N)$.

\subsubsection{Time Complexity for Maintaining Segments}
In this section, we use amortized analysis to compute the total cost for maintaining segments. Consider an arbitrary segment $E$. Whenever a node containing $E$ is visited during the recursion, a subset of the following operations are performed:

\begin{itemize}
    \item Adding $E$ into $P.segments$ when first created.
    \item Removing $E$ from $P.segments$. If each node stores segments in a linked list and each segment keeps track of its location inside the list, this operation can be done in $O(1)$ time.
    \item Modifying $E.back$ and $E.front$.
    \item Splitting $E$ into two new segments $E_1$ and $E_2$. This is equivalent to first changing $E$ to $E_1$, then creating $E_2$.
\end{itemize}

Therefore, the cost for visiting $E$ is $O(1)$. When $E$ is created, it points to two nodes in BVH. When $E$ is visited, at least one of $E$'s pointer must move downward in BVH. $E$ is never destroyed. This indicates that the number of visits to $E$ is bounded by the height of BVH, which is $O(\log N)$. Combined with previous analysis, each segment costs $O(\log N)$ time.

Assume $M$ is the total number of segments. Then all segments point to $2M$ blocks. Since there are $O(\frac{N}{B})$ blocks, M is bounded by $O(\frac{N}{B})$. Thus, the total time for maintaining sefments is $O(\frac{N}{B} \log N)$.

\subsubsection{Space Complexity}
Constructing BVH can be implemented as in-place computation, where all points are stored in a global array. Each node points to a continuous interval of the global array, and sorts this interval during the recursion. This method not only reduces memory usage to $O(N)$, but also increases spatial locality of the program. Segments occupy $O(M)$ space. BVH has the structure of a perfect binary search tree, which consumes $O(N)$ space. The total time complexity is thus $O(N)$.

\section{Experiments}
\subsection{Benchmark Problems}
In order to benchmark augmented BVH, we created random graphs with degrees ranging from 50 to 1000 (step size = 50). For each degree, all algorithms are tested in 10 random graphs. This results in total 200 graphs.

\subsection{Experiment Setup}
% three-step algorithm
\begin{figure}[htbp]
    \centering
    \begin{subfigure}[b]{0.32\textwidth}
        \includegraphics[width=\textwidth]{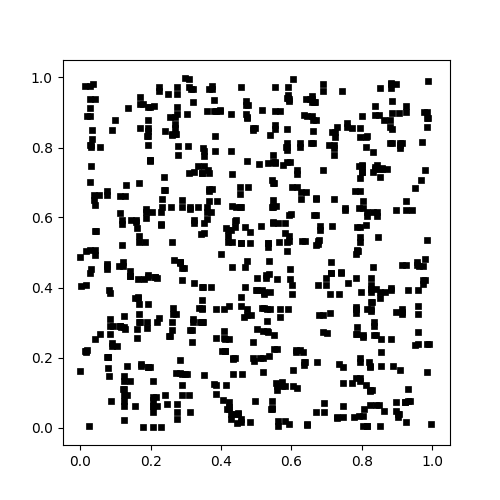}
        \caption{Input Graph}
        \label{fig:step1}
    \end{subfigure}
    \hfill
    \begin{subfigure}[b]{0.32\textwidth}
        \includegraphics[width=\textwidth]{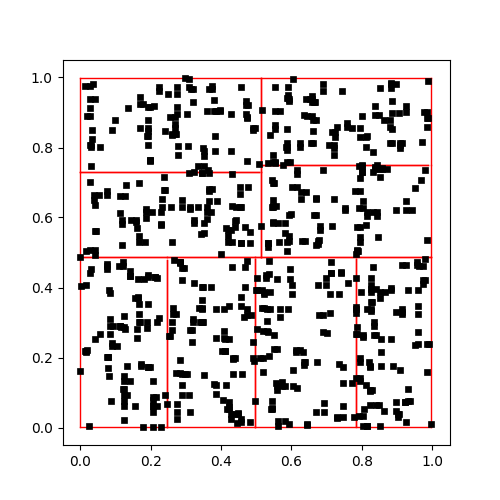}
        \caption{Divide}
        \label{fig:step2}
    \end{subfigure}
    \hfill
    \begin{subfigure}[b]{0.32\textwidth}
        \includegraphics[width=\textwidth]{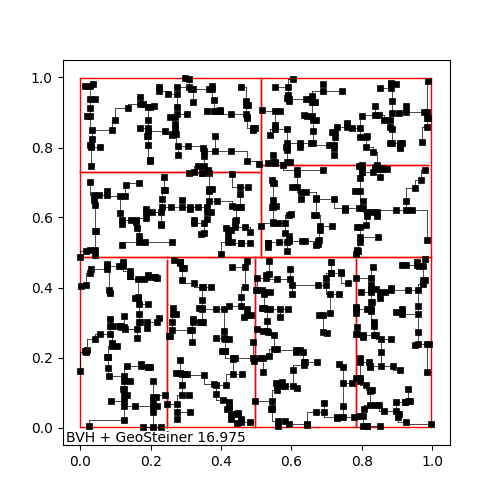}
        \caption{Conquer}
        \label{fig:step3}
    \end{subfigure}
    \hfill
    \caption{Steps}
    \label{fig:steps}
\end{figure}

The full procedure of our algorithm is shown in \cref{fig:steps}. Assume $\Phi$ is the time complexity for the chosen solver. The cost for constructing BVH is $O(N \log N)$. Computing local Steiner trees takes $O\left(\frac{N}{B} \times \Phi(B)\right)$ time. To compute minimum spanning tree, we first conpute distances betten subtrees, which takes $O\left(\frac{N}{B} \times B^2\right)$ time. Finally, Prim's algorithm \cite{Prim1957} takes $O\left(\frac{N}{B} \log \frac{N}{B} \right)$ time. The total time complexity is

$$
O\left(N \times \left( \frac{\Phi(B)}{B} + B + \log N \right) \right)
$$

In our experiment, we used single process for all algorithms to compare asymptotic complexities. However, it is worth noticing that our algorithm allows concurrency. Constructing the left and right child of a BVH node is independent, so the recursion process can be parallelized. Local Steiner trees are independent with each other, allowing further parallelization. 

We chose GeoSteiner, REST and FLUTE as candidate algorithms. They each represent a special class of RSMT solver. $B$ was set to 100 for GeoSteiner and FLUTE. Since our REST agent was trained on graphs with 50 points, $B$ was set to 50 for REST. Accuracy constant $A$ in FLUTE was set to 10.

\subsection{Comparison}
\begin{figure}[htbp]
    \centering
    \begin{subfigure}[b]{0.32\textwidth}
        \includegraphics[width=\textwidth]{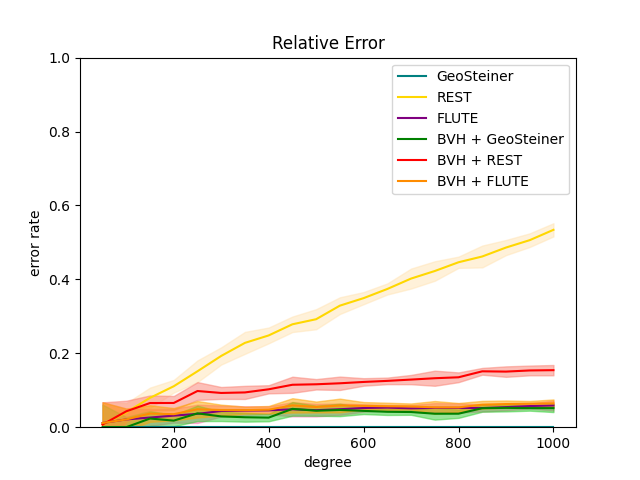}
        \caption{Error Rate}
        \label{fig:time}
    \end{subfigure}
    \hfill
    \begin{subfigure}[b]{0.32\textwidth}
        \includegraphics[width=\textwidth]{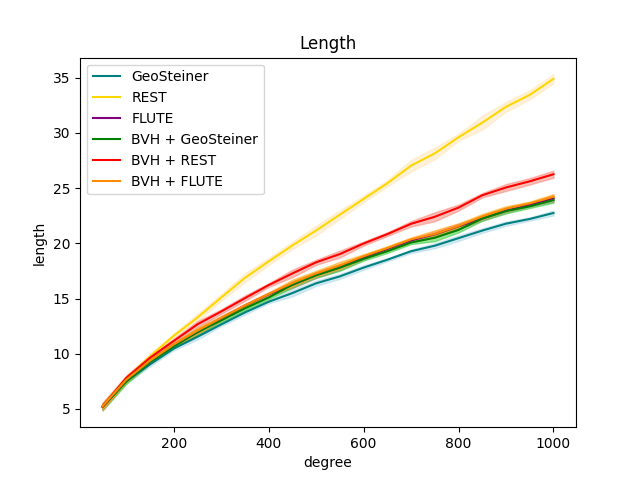}
        \caption{Length}
        \label{fig:length}
    \end{subfigure}
    \hfill
    \begin{subfigure}[b]{0.32\textwidth}
        \includegraphics[width=\textwidth]{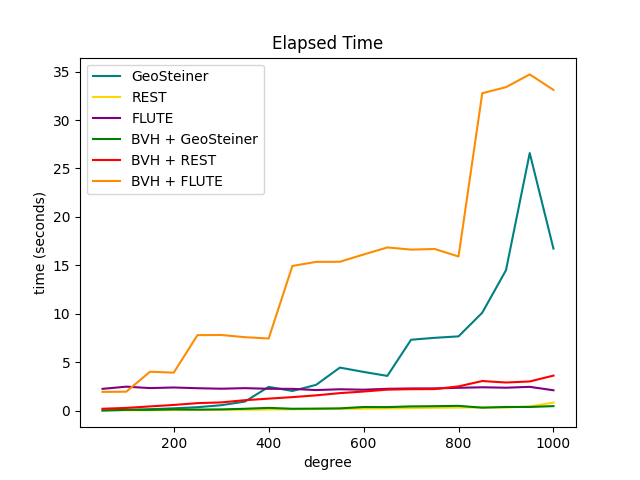}
        \caption{Time}
        \label{fig:error}
    \end{subfigure}
    \caption{Results}
    \label{fig:results}
\end{figure}

Running time, routing length and relative error are summarized in \cref{tab:time,tab:length}. To avoid confusion, "BVH+G", "BVH+R" and "BVH+F" represent applying augmented BVH to GeoSteiner, REST and FLUTE respectively. Results are visualized in \cref{fig:results}. To evaluate the effectiveness of our approach, we compare GeoSteiner, REST, and FLUTE against their augmented counterparts. We analyze the performance from both vertical and horizontal perspectives.

% figures

% time
\begin{longtable}{c|*{6}{c}}
    \caption{Time (seconds)} \label{tab:time} \\
    \toprule
    \cmidrule(lr){2-7}
    degree & GeoSteiner & REST & FLUTE & BVH+G & BVH+R & BVH+F \\
    \midrule
    \endfirsthead
    \toprule
    \cmidrule(lr){2-7}
    degree & GeoSteiner & REST & FLUTE & BVH+G & BVH+R & BVH+F \\
    \midrule
    \endhead
    \midrule
    \endfoot
    \bottomrule
    \endlastfoot 

    50 & 0.01 & \textbf{0.06} & 2.24 & 0.01 & 0.18 & 1.93 \\
    100 & 0.06 & \textbf{0.03} & 2.47 & 0.07 & 0.27 & 1.95 \\
    150 & 0.14 & \textbf{0.03} & 2.32 & 0.07 & 0.43 & 4.01 \\
    200 & 0.23 & \textbf{0.06} & 2.38 & 0.12 & 0.58 & 3.91 \\
    250 & 0.34 & \textbf{0.06} & 2.31 & 0.10 & 0.76 & 7.79 \\
    300 & 0.55 & \textbf{0.07} & 2.25 & 0.12 & 0.84 & 7.80 \\
    350 & 0.93 & \textbf{0.09} & 2.31 & 0.18 & 1.07 & 7.57 \\
    400 & 2.44 & \textbf{0.11} & 2.25 & 0.27 & 1.23 & 7.44 \\
    450 & 2.02 & \textbf{0.13} & 2.24 & 0.18 & 1.39 & 14.93 \\
    500 & 2.66 & \textbf{0.16} & 2.11 & 0.20 & 1.57 & 15.35 \\
    550 & 4.43 & \textbf{0.17} & 2.20 & 0.22 & 1.80 & 15.36 \\
    600 & 3.99 & \textbf{0.20} & 2.15 & 0.36 & 1.96 & 16.12 \\
    650 & 3.58 & \textbf{0.22} & 2.24 & 0.35 & 2.16 & 16.84 \\
    700 & 7.31 & \textbf{0.25} & 2.28 & 0.43 & 2.21 & 16.62 \\
    750 & 7.51 & \textbf{0.29} & 2.29 & 0.45 & 2.22 & 16.68 \\
    800 & 7.66 & \textbf{0.30} & 2.35 & 0.48 & 2.49 & 15.91 \\
    850 & 10.10 & 0.34 & 2.40 & \textbf{0.29} & 3.05 & 32.77 \\
    900 & 14.49 & \textbf{0.31} & 2.36 & 0.37 & 2.89 & 33.40 \\
    950 & 26.59 & 0.42 & 2.45 & \textbf{0.37} & 3.00 & 34.72 \\
    1000 & 16.72 & 0.82 & 2.09 & \textbf{0.46} & 3.61 & 33.11 \\
\end{longtable}

% length
\begin{longtable}{c|*{6}{c}}
    \caption{Length and Relative Error} \label{tab:length} \\ 
    \toprule
    \cmidrule(lr){2-7}
    degree & GeoSteiner & REST & FLUTE & BVH+G & BVH+R & BVH+F \\
    \midrule
    \endfirsthead 
    \toprule
    \cmidrule(lr){2-7}
    degree & GeoSteiner & REST & FLUTE & BVH+G & BVH+R & BVH+F \\
    \midrule
    \endhead 
    \midrule
    \endfoot 
    \midrule
    \multicolumn{7}{r}{{Continued on next page}} \\ 
    \endfoot
    \bottomrule
    \endlastfoot 

    50 & 5.19 & 5.22 (0.71) & 5.25 (1.14) & 5.19 (0.00) & 5.22 (0.71) & 5.25 (1.14) \\
    100 & 7.49 & 7.80 (4.03) & 7.65 (2.10) & 7.49 (0.00) & 7.82 (4.30) & 7.65 (2.10) \\
    150 & 9.04 & 9.76 (7.94) & 9.28 (2.61) & 9.26 (2.35) & 9.63 (6.54) & 9.38 (3.69) \\
    200 & 10.47 & 11.63 (11.08) & 10.80 (3.13) & 10.66 (1.78) & 11.16 (6.56) & 10.84 (3.52) \\
    250 & 11.53 & 13.28 (15.13) & 11.95 (3.61) & 11.97 (3.74) & 12.66 (9.79) & 12.09 (4.85) \\
    300 & 12.66 & 15.10 (19.28) & 13.21 (4.39) & 13.02 (2.92) & 13.83 (9.27) & 13.24 (4.63) \\
    350 & 13.74 & 16.88 (22.82) & 14.36 (4.49) & 14.11 (2.69) & 15.03 (9.41) & 14.38 (4.62) \\
    400 & 14.71 & 18.36 (24.83) & 15.38 (4.56) & 15.09 (2.59) & 16.22 (10.26) & 15.41 (4.74) \\
    450 & 15.49 & 19.81 (27.84) & 16.25 (4.85) & 16.26 (4.96) & 17.27 (11.48) & 16.41 (5.90) \\
    500 & 16.38 & 21.16 (29.22) & 17.13 (4.61) & 17.11 (4.46) & 18.28 (11.61) & 17.27 (5.47) \\
    550 & 17.01 & 22.60 (32.89) & 17.83 (4.86) & 17.80 (4.67) & 19.03 (11.88) & 18.01 (5.88) \\
    600 & 17.80 & 24.02 (34.92) & 18.73 (5.20) & 18.59 (4.42) & 19.98 (12.24) & 18.83 (5.77) \\
    650 & 18.52 & 25.44 (37.40) & 19.49 (5.29) & 19.29 (4.17) & 20.84 (12.53) & 19.57 (5.68) \\
    700 & 19.28 & 27.04 (40.21) & 20.27 (5.11) & 20.07 (4.11) & 21.77 (12.88) & 20.36 (5.60) \\
    750 & 19.78 & 28.14 (42.26) & 20.82 (5.22) & 20.51 (3.65) & 22.41 (13.25) & 20.85 (5.37) \\
    800 & 20.47 & 29.60 (44.63) & 21.54 (5.22) & 21.22 (3.66) & 23.23 (13.51) & 21.57 (5.37) \\
    850 & 21.15 & 30.93 (46.19) & 22.25 (5.18) & 22.24 (5.14) & 24.35 (15.12) & 22.44 (6.09) \\
    900 & 21.78 & 32.37 (48.61) & 22.96 (5.44) & 22.92 (5.23) & 25.06 (15.04) & 23.14 (6.26) \\
    950 & 22.21 & 33.44 (50.61) & 23.48 (5.72) & 23.35 (5.16) & 25.62 (15.35) & 23.62 (6.35) \\
    1000 & 22.75 & 34.89 (53.40) & 24.08 (5.88) & 23.92 (5.16) & 26.26 (15.43) & 24.21 (6.44) \\
\end{longtable}

\subsubsection{Vertical Comparison}
Among the three baseline algorithms, FLUTE maintains the fastest execution time across all tested degrees. Its performance advantage slightly diminishes as the problem size increases. REST, while being computationally efficient, exhibits a rapid increase in relative error as the degree grows beyond 500. GeoSteiner the highest time complexity among the three baseline algorithms. 

The introduction of augmented BVH significantly improves the performance of GeoSteiner. The runtime of GeoSteiner is reduced by an order of magnitude, particularly for instances with more than 500 terminals, making it competitive with FLUTE in terms of runtime while maintaining higher accuracy. 

The error rate of REST is significantly reduced while maintaining relatively high speed for large graphs. For graphs with 1000 terminals, the relative error drops from 53.4\% to 15.4\%. The asymptotic time complexity of augmented REST is dominated by the $O\left(\frac{N}{B} \times \Phi(B)\right)$ term. In our experiment, we constructed an REST agent for each subgraph and computed all subtrees sequentially. Frequent memory transactions between CPU and GPU lead to high $O(\Phi(B))$. This can be optimized by batched operations, where we compute subtrees in parallel.

Large time constant becomes obvious in FLUTE. Although the accuracy of FLUTE is maintained, our algorithm suffers from high cost. As shown in \cref{tab:time}, FLUTE consumes at least 2 seconds regardless of input degree. This indicates the potential weakness of our algorithm: $O(\Phi(B))$ should be monotonically increasing and passes through the origin. These issues can be overcame with concurrency.

\subsubsection{Horizontal Comparison}
% Pareto
\begin{figure}[ht]
    \centering
    \includegraphics[width=0.48\textwidth]{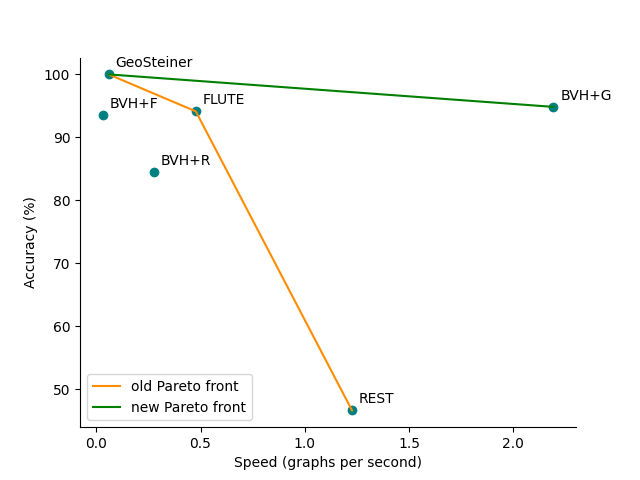}
    \caption{Pareto Front}
    \label{fig:Pareto}
\end{figure}

Designing and implementing RSMT solvers can be seen as a multi-objective optimization procedure. In the case of RSMT, the two objectives are accuracy and speed. No single optimal algorithm exists that maximizes all objectives. Instead, we achieve a set of non-dominated algorithms called Pareto set. The image of Pareto set in objective space leads to the notion of Pareto front. The quality of Pareto front can be measured by hypervolume metric \cite{zitzler1999multiobjective} and sparsity metric.

\cref{fig:Pareto} visualizes all algorithms in objective space. Previously, GeoSteiner, REST and FLUTE form a Pareto front. After boosted by algorithm, GeoSteiner and augmented GeoSteiner form the new Pareto front. The new Pareto front has significantly better hypervolume metric. However, the current Pareto front has a weak sparsity metric. By boosting more RSMT solvers with parallelized version of our method, the Pareto front could be further extended.

In summary, augmented GeoSteiner provides the best overall accuracy and runtime, making it a strong candidate for large-scale RSMT problems. Augmented BVH transforms REST into a more practical solver by improving its accuracy while maintaining its efficiency. Our method ensures that FLUTE remains highly applicable, but suffers from unnecessary complexity. Augmented BVH provides a new class of solvers out of existing ones, pushing the Pareto front forward in RSMT computation.

\section{Conclusion}
In this work, we propose a BVH based algorithm that boosts any RSMT solver. Our algorithm produces competitive results for medium to large size graphs in terms of speed and accuracy. This provides a way to utilize various solvers in reality for different needs.

\bibliographystyle{unsrt}
\bibliography{papers}

\end{document}